\documentclass[]{interact}

\usepackage{epstopdf}
\usepackage[caption=false]{subfig}
\usepackage{graphicx, rotating}
\usepackage{mathptmx}      
\usepackage{latexsym}
\usepackage{amssymb}
\usepackage{mathtools}
\usepackage{bm}
\usepackage{hyperref}
\usepackage{cleveref}
\usepackage[flushleft]{threeparttable}
\usepackage[authoryear]{natbib}
\bibpunct[, ]{(}{)}{;}{a}{,}{,}

\theoremstyle{plain}

\theoremstyle{definition}

\theoremstyle{remark}

\DeclareMathOperator*{\argmin}{arg\,min}
\begin{document}


\title{PORTFOLIO OPTIMIZATION FOR INDEX TRACKING WITH CONSTRAINTS ON DOWNSIDE RISK AND CARBON FOOTPRINT}

\author{
\name{Suparna Biswas\thanks{CONTACT Suparna Biswas. Email: suparnabsws4@gmail.com} and Rituparna Sen}
\affil{Indian Statistical Institute, Bangalore, Karnataka, 560059, India}
}

\maketitle

\begin{abstract}
Historically, financial risk management has mostly addressed risk factors that arise from the financial environment. Climate risks present a novel and significant challenge for companies and financial markets. Investors aiming for avoidance of firms with high carbon footprints require suitable risk measures and portfolio management strategies. This paper presents the construction of decarbonized indices for tracking the S \& P-500 index of the U.S. stock market, as well as the Indian index NIFTY-50, employing two distinct methodologies and study their performances. These decarbonized indices optimize the portfolio weights by minimizing the mean-VaR and mean-ES and seek to reduce the risk of significant financial losses while still pursuing decarbonization goals. Investors can thereby find a balance between financial performance and environmental responsibilities. Ensuring transparency in the development of these indices will encourage the excluded and under-weighted asset companies to lower their carbon footprints through appropriate action plans. For long-term passive investors, these indices may present a more favourable option than green stocks.\

\end{abstract}


\section{Introduction}\label{sec1}

In the past, financial risk management has mostly dealt with risks that come from the financial world. Nevertheless, non-financial sources of risk have arisen as crucial determinants for the functioning of enterprises and organizations. Climate risks have garnered considerable attention from policymakers, institutions, and investors worldwide. This is due to the growing recognition of the potentially dire consequences of climate change for the financial system, as well as the integration of these risks into decision-making processes (see \cite{TCFD22}). Climate risks, encompassing both transition and physical risks, present a novel and significant challenge for companies and financial markets (see \cite{breitenstein2022}). Extreme weather events like floods and heat waves, as well as changes in climatic patterns like droughts and rising sea levels, are examples of physical risks brought on by climate change that can directly harm organizations and assets. Transition risk refers to the potential financial losses that can arise from the shift towards a low-carbon economy, driven by factors like policy changes, technological advancements, and evolving consumer preferences. Integrating these risks into forecasting and planning processes introduces new complexities (see \cite{capelli2023}). Also, climate concerns are not limited to the financial sector and are global in nature. Anecdotal evidence suggests that there is a strong market demand for climate management (see \cite{clark2023}). In the past few years, European financial authorities have pushed for investment companies to better include environmental, social, and governance (ESG) risks in their risk management and governance frameworks (\cite{EBA21}). Certain researchers propose that to mitigate environmental impacts, consumption patterns must transition towards cleaner products \cite{schandl16}, while others argue that environmental degradation is inextricably linked to economic growth, necessitating substantial transformations in economies to render degrowth viable \cite{kallis2011}. Both perspectives indicate that financial systems must undergo significant adjustments to align with the new paradigm, necessitating that investors identify suitable strategies for sustainable investment to mitigate their risks while facilitating the economy's transition from unsustainable production methods.\

Investors and financial institutions must therefore engage in the global push to decarbonize economies. For investors to accurately assess which organizations to invest in, relevant indicators and measures must exist that reflect both the qualitative and quantitative effects of firms' operations on the environment. When appropriate measures are implemented, investors necessitate strategies that can optimize their portfolios in order to mitigate the risk estimated by the risk measures. Markowitz \cite{Mark1952} introduced the optimal portfolio choice theory in 1952, which is based on the risk of variance. Since then, the variance (mean variance) has become a classic and highly influential quantitative measure of financial risks. The computation of variance is straightforward and user-friendly, and the underlying theory has been extensively advanced. Nevertheless, it only takes into account the average deviation, failing to address the prevalent concern of the right tail of the returns.  It is desirable for measures of risk to concentrate on the tail of
the distribution of returns, in other words, the extreme losses. Downside risk refers to the potential decrease in an asset's value due to fluctuations in market conditions. The measures of downside risk, like value-at-risk (VaR) and expected shortfall (ES), are adopted as standard tools to measure the ex-ante financial risk of assets (BCBS, 1996 and Basel III). VaR quantifies the size of loss on a portfolio of assets over a given time horizon at a given probability. ES measures the weighted average of the ``extreme'' loss beyond the VaR cutoff point. The ES is a coherent risk measure, whereas VaR is not. The VaR method was developed to measure financial market risk in response to the financial disasters of the early 1990s and to address the demerits of VaR, ES was introduced by \cite{artzner1999}. \cite{capelli21} saw ESG factors as one of the parts that are missing when VaR is used to measure financial risk.\

This research presents methodologies for constructing two decarbonized indices from established benchmarks (see \cite{lakshmi23}), and demonstrates their effectiveness for the Indian and U.S. economies. We demonstrate that the resultant index substantially reduces overall carbon effect, serving as a safeguard against climate hazards. Our approach is based on VaR and ES. The relationship of VaR and ES to ESG has remained largely unexplored in portfolio optimization. In a pertinent work by \citep{andersson2016hedging}, the presented decarbonized indices were derived from the benchmark by minimizing tracking error while adhering to appropriate limitations related to the carbon footprints of the constituent enterprises. \citep{mezali2013} earlier used quantile regression with a mixed-integer linear programming formulation. \citep{li2022robust} developed a comprehensive methodology that optimizes ESG score while concurrently minimizing risk and maximizing revenue.\

It is essential to recognize that the current sustainability-oriented indices in the Indian and U.S. stock markets, specifically the S \& P BSE GREENEX, BSE Carbonex, NIFTY100 Enhanced ESG Index \citep{patel2020indian,nishad2021carbon}, and S \& P-500 Scored \& Screened Index \cite{SNP}, emphasize monitoring the performance of companies based on their carbon emissions, ESG scores, and initiatives to mitigate climate risk, rather than concentrating on the performance of the parent index. They employ market capitalization for weighting, without attempting to imitate the performance of the excluded equities. To circumvent this, \citep{andersson2016hedging,lakshmi23} develop an optimized index by minimizing tracking error. To determine how closely a portfolio tracks its benchmark index, tracking error is a crucial tool. Tracking errors, however, don't provide any indication of the DIs' risk. Here, we develop an optimized index by minimizing the mean-VaR and mean-ES. It effectively captures lost contributions from dropped stocks by compensating with other highly correlated stocks that remain in the portfolio, while minimizing VaR and ES.\

We state the definition and describe the methodology in Section 2. Section 3 provides the data description. The application of the methods on the Indian and the U.S. markets is illustrated in Section 4. We demonstrate how the index aims to establish a connection between theory and practice by using real-time data for both in-sample and out-of-sample computations. The paper ends with a detailed summary and discussion in Section 5.

\section{Definition and Methodology}

In this section we define VaR and ES. We then discuss the methodologies behind the construction of decarbonized indices and then optimize the portfolio weights by minimizing the mean-VaR and mean-ES.
\subsection{Definitions}
\noindent Let the random variable $X$ be the loss of some portfolio. Let $F$ be the distribution function of $X$, then $Q_p(X)=\inf\{x:F(x)\geq p\},\ 0<p<1$ is the quantile function. For, $0<p<q<1$, the two risk measures $VaR_p$ and $ES_p$ are defined as
$$VaR_p=\inf\{x\in\mathbb{R}:F(x)\geq p\}$$ and
\begin{equation}\label{ES}ES_p=\frac{1}{1-p}\int_p^1VaR_udu.\end{equation}

\subsection{Methodology}
Let $N$ stocks be arranged from highest to lowest based on their carbon footprints. For the $i$th stock $r_i$, $M_i$, and $c_i$ denote the return, market capitalization, and carbon footprint, respectively. The portfolio return of the benchmark is $R^b=(\textbf{W}^b)^T\textbf{r}$, where $\textbf{W}^b=(W_i^b)_{1\leq i\leq N}$ is the vector of portfolio weights taken to be proportional to the market capitalization,
\begin{equation}\label{wb}
W_i^b=\frac{M_i}{\sum_{i=1}^NM_i},
\end{equation}
and the bold letters denote the corresponding vector representation. Let $\textbf{W}^d$ be the vector of weights for the proposed decarbonized index and $R^d$ is the corresponding return. We assume that the return rate $\textbf{r}\sim N(\mu,\Sigma)$, then
$$R^b\sim N((\textbf{W}^b)^T\mu,(\textbf{W}^b)^T\Sigma\textbf{W}^b)$$ and $$R^d\sim N((\textbf{W}^d)^T\mu, (\textbf{W}^d)^T\Sigma\textbf{W}^d).$$

Our objective is to minimize the mean-VaR and mean-ES and then find
\begin{eqnarray}\label{VaRD}
\textbf{W}^d&=&\argmin_{\textbf{W}=(W_i)_{1 \le i \le N}} (VaR_p(\textbf{W}^T\textbf{r})) \\ \notag
&=&\argmin_{\textbf{W}=(W_i)_{1 \le i \le N}}(\textbf{W}^T\mu+T(p)\sqrt{\textbf{W}^T\Sigma\textbf{W}}),
\end{eqnarray}
$T(p)$ is the $p$th quantile of the standard normal distribution, and
\begin{eqnarray}\label{ESD}
\textbf{W}^d&=&\argmin_{\textbf{W}=(W_i)_{1 \le i \le N}} (ES_p(\textbf{W}^T\textbf{r})) \\ \notag
&=&\argmin_{\textbf{W}=(W_i)_{1 \le i \le N}}(\textbf{W}^T\mu+T_1(p)\sqrt{\textbf{W}^T\Sigma\textbf{W}}),
\end{eqnarray}
$T_1(p)=\frac{\phi(\Phi^{-1}(p))}{1-p}$, where $\phi$ is the probability density function of the standard normal distribution and $\Phi$ is the cumulative distribution function of the standard normal distribution.

In order to reduce the computations, we use a multi-factor model of aggregate risk to estimate VaR and ES. Next for S \& P-500 index, we use the \cite{fama2012} five-factor model and for the Indian benchmark index Nifty-50, we use the \cite{fama2012} four-factor model, which allows us to decompose the return into a weighted sum of common factor returns and specific returns. If $r_{it}$ and $r_{ft}$ denote the return of the $i$th stock and the risk-free rate at time $t$, then the five-factor model is,
\begin{equation}\label{FM5}
    r_{it} - r_{ft} = \beta_{i0} + \beta_{i1} (r_{mt}-r_{ft}) + \beta_{i2} \mathrm{SMB}_t + \beta_{i3}\mathrm{HML}_t + \beta_{i4} \mathrm{RMW}_t + \beta_{i5}\mathrm{CMA}_t+ e_{it},
\end{equation}
and the four-factor model is,
\begin{equation}\label{FM4}
    r_{it} - r_{ft} = \beta_{i0} + \beta_{i1} \mathrm{SMB}_t + \beta_{i2} \mathrm{HML}_t + \beta_{i3}\mathrm{WML}_t + \beta_{i4} \mathrm{MF}_t + e_{it},
\end{equation}
where $e_{it}$ is the error, $\beta_{ij}$ denotes the factor loading, and $r_{mt}-r_{ft}$ is the market risk premium; $\mathrm{SMB}$, $\mathrm{HML}$, $\mathrm{WML}$, $\mathrm{MF}$, $\mathrm{RMW}$, and $\mathrm{CMA}$, indicate the size effect (small-minus-big), value effect (high-minus-low), momentum factor (winners-minus-losers), market factor, the return spread of the most profitable firms minus the least profitable, and the return spread of firms that invest conservatively minus aggressively. Let $F_j$ denote these factors with a dispersion matrix $\Omega$. Also, let $\bm\beta$ be the matrix of loadings and $\Delta$ be the diagonal matrix of specific risk variances. Then, the dispersion of the excess returns is $\bm\beta \Omega \bm\beta^T + \Delta$. Consequently, the volatility of any portfolio with returns \textbf{$r$} and weights $\textbf{w}$ is $\sqrt{\textbf{w}^T(\bm\beta \Omega \bm\beta^T + \Delta) \textbf{w}}$. Now, equation \ref{VaRD} and \ref{ESD} can be written as

\begin{eqnarray}\label{VaRD1}
\textbf{W}^d&=&\argmin_{\textbf{W}=(W_i)_{1 \le i \le N}}(\textbf{W}^T\mu+T(p)\sqrt{\textbf{W}^T(\bm\beta \Omega \bm\beta^T + \Delta)\textbf{W}}),
\end{eqnarray}
and
\begin{eqnarray}\label{ESD1}
\textbf{W}^d&=&\argmin_{\textbf{W}=(W_i)_{1 \le i \le N}}(\textbf{W}^T\mu+T_1(p)\sqrt{\textbf{W}^T(\bm\beta \Omega \bm\beta^T + \Delta)\textbf{W}}),
\end{eqnarray}
In order to achieve a balance between minimizing carbon footprints and maintaining diversity in composition, we utilize two separate methodologies to create decarbonized indices (DIs). These two methodologies have been taken from \citep{andersson2016hedging,lakshmi23}. Every methodology presents unique benefits and drawbacks, which are detailed below.

The first method involves reweighting the remaining stocks in order to reduce mean-VaR and mean-ES after excluding the $k$ worst performers in terms of carbon intensity. Here, the DI is constructed using weights $W^d_i$, obtained by solving \eqref{VaRD1} and \eqref{ESD1} subject to the constraints

\begin{equation}\label{method1}
    \begin{split}
        & \sum_{i=1}^{N} W_i^d = 1, \; \text{with} \\
        & W_i^d = 0, \; \text{for } i= 1,2,\ldots,k, \; \text{and} \; 0 \le W_i^d \le 1, \; \text{for } i=k+1,\dots,N.
    \end{split}
\end{equation}

The minimization problem is addressed using the Trust-Region Constrained Algorithm (TRCA), which is effective for handling the following problem:
\begin{equation}\label{minimization-problem-1}
	\text{minimize} \; f(x), \text{ subject to } c_0^{lb} \le c_0(x) \le c_0^{ub}, \;  x^{lb} \le x \le x^{ub}.
\end{equation}

Multiple linear and nonlinear constraints can be accepted as inputs \citep{conn2000trust}. The objective function is estimated using a quadratic model confined to the trust region centered on the initial estimate or the present location. The algorithm operates through a process of iterative refinement of the initial estimate \citep{kimiaei2022active}. We exclude the algorithm's technicalities and direct the reader to \cite{byrd1987trust} for additional information.

Our second methodology includes all stocks without specifically targeting those with high carbon footprints. The minimization problem outlined in \eqref{VaRD1} and \eqref{ESD1} is addressed by establishing a threshold $C$ for the total footprint of the index. This method preserves the composition's diversity while minimizing its footprint. Mathematically, we find the weights in \eqref{VaRD1} and \ref{ESD1} considering
\begin{equation}\label{method2}
    \begin{split}
        & \sum_{i=1}^{N} W_i^d = 1, \; \text{with} \\
        & \sum_{i=1}^{N} c_iW_i^d \le C \; \text{and} \; 0 \le W_i^d \le 1, \; \text{for } i=1,\hdots,N.
    \end{split}
\end{equation}

We now implement the TRCA mentioned above. Hereafter, the first DI index is denoted as DI\_1 and the second index is denoted as DI\_2.

In this context, it is essential to conduct a quick comparison of the ideologies that were used in the building of the indices.  One of the potential drawbacks of the first technique is that it might result in a composition of the index that is less diversified.  A lower level of diversity results in increased risk and volatility.  On the bright side, the prospect of inclusion in the index can perform the function of an incentive for high-emission enterprises to cut their emissions in a proactive manner.  When compared to the first strategy, the second approach results in a significant reduction in the overall carbon footprint; however, this reduction is not as great as the first approach.

\section{Data}

The closing price data for the constituent stocks of S \& P-500 and NIFTY-50 were obtained from Yahoo! Finance using the Python yfinance module. Similarly, the market cap data for each stock was downloaded using the Python pandas-datareader from Yahoo.

\subsection{Carbon Data}

The carbon emissions data for U.S. equities and Indian equities were obtained from the Bloomberg Terminal of the Indian Institute of Management, Bangalore (IIM-B). It included 14 different emissions measures. We chose the two variables greenhouse gas intensity per sale (GHG) and total carbon dioxide emissions (CO2) as proxies for the carbon footprint of stocks. These variables have lesser missing values for five years from 2017-2022.
The factor data for \eqref{FM4} are obtained from the IIM-A Data Library \citep{AJV2013-India-4Factor} and for \eqref{FM5} it is available at \url{https://mba.tuck.dartmouth.edu/pages/faculty/ken.french/data_library.html}. The Fama-French five-factor data used in the regression step for the S \& P-500 index
was obtained from Kenneth R. French - Data Library. Weekly and daily data are available in the data library, of which we have used the daily data.

\section{Investment Portfolio Based on VaR and ES}

We consider NIFTY-50 and S \& P-500 data for 5 years, from 2017-18 until 2022-23. Then, four DIs are created from each benchmark, using the two methods and the two proxies, GHG and CO2. Comprehensive information about the stocks used for our calculations is reported in \Cref{NSE} and \Cref{SNP}.\\

\begin{table}[!ht]
    \centering
    \caption{Number of stocks included (SI), stocks omitted (SO), and corresponding omission percentage of market capitalization MCO in the construction of DI for NIFTY-50.}\label{NSE}
    \begin{tabular}{|l|ccc|ccc|}
    \hline
    & \multicolumn{3}{|c|}{GHG} & \multicolumn{3}{|c|}{CO2} \\
    Period & SI & SO & MCO & SI & SO & MCO \\
    \hline
    2017-18 & 30 & 20 & 35.3\% & 32 & 18 & 33.7\% \\
    2018-19 & 33 & 17 & 28.5\% & 35 & 15 & 26.9\% \\
    2019-20 & 34 & 16 & 25.3\% & 36 & 14 & 23.7\% \\
    2020-21 & 35 & 15 & 23.6\% & 38 & 12 & 21.1\% \\
    2021-22 & 35 & 15 & 23.6\% & 38 & 12 & 21.0\% \\
    \hline
    \end{tabular}
\end{table}

\begin{table}[!ht]
    \centering
    \caption{Number of SI, SO and corresponding MCO in the construction of DI for S \& P-500.}\label{SNP}
    \begin{tabular}{|l|ccc|ccc|}
    \hline
    & \multicolumn{3}{|c|}{GHG} & \multicolumn{3}{|c|}{CO2} \\
    Period & SI & SO & MCO & SI & SO & MCO \\
    \hline
    2017-18 & 331 & 171 & 21.0\% & 335 & 167 & 20.6\% \\
    2018-19 & 366 & 136 & 17.4\% & 368 & 134 & 17.7\% \\
    2019-20 & 408 & 94 & 13.9\% & 410 & 92 & 13.7\% \\
    2020-21 & 432& 70 & 9.3 & 434 & 68 & 10.8\% \\
    2021-22 & 435 & 67 & 9.1\% & 436 & 66 & 9.57\% \\
    \hline
    \end{tabular}
\end{table}

Our analysis is divided into three main components: first, we will determine the optimal values of $k$ and $C$ for calculating the two DIs for each proxy and each benchmark; second, we will generate optimal portfolio weights using a one year window over a five year period (in-sample calculations); and third, we will calculate the monthly performance of the DIs and compare their performances against the benchmark (out-of-sample calculations).

The optimal values of $k$ and $C$ (refer to \eqref{method1} and \eqref{method2}), are ascertained by evaluating VaR and ES utilizing five years of data. A series of optimizations are conducted for various values of $k$ (5\%-50\% of $N$) and $C$ (50\%-95\%). We plot $k$ vs VaR and $k$ vs ES and then find the value $k$ for which VaR and ES are minimum. Similarly, we choose the optimal carbon footprint threshold by plotting the relation between the percentage of carbon footprint and the VaR and ES of the portfolio and choose the lowest value that considerably reduces the VaR and ES in each case. The plots are not shown here; instead, the obtained optimal values of $k$ and $C$ are reported in \Cref{OP}.

\begin{table}[!ht]
    \centering
    \caption{Optimal values of $k$ and $C$ estimated for each index}\label{OP}
    \begin{tabular}{|l|cccc|cccc|}
    \hline
    &\multicolumn{4}{|c|}{VaR} & \multicolumn{4}{|c|}{ES} \\
    & \multicolumn{2}{|c|}{GHG} & \multicolumn{2}{|c|}{CO2} & \multicolumn{2}{|c|}{GHG} & \multicolumn{2}{|c|}{CO2} \\
    Period & k & C &   k & C & k & C &   k & C \\
    \hline
    NIFTY-50 & 1 & 95\%  & 2 & 95\% & 1 & 95\%  & 2 & 95\% \\
    S \& P-500 & 48 & 85\%  & 32 & 95\% & 16 & 95\%  & 32 & 95\% \\
    \hline
    \end{tabular}
\end{table}
Once the optimal $k$ and $C$ are obtained, the next step is to construct the DIs and determine the portfolio weights. We use a moving window of one year from April 2017 to March 2022 for in-sample estimations. In the next two sections we discuss the DIs obtained through VaR and ES assessment and our main findings from the in-sample estimation.

\subsection{DIs through VaR assessment}

\begin{figure}[h]
\centering
  \centering
  \includegraphics[width=0.8\textwidth,keepaspectratio]{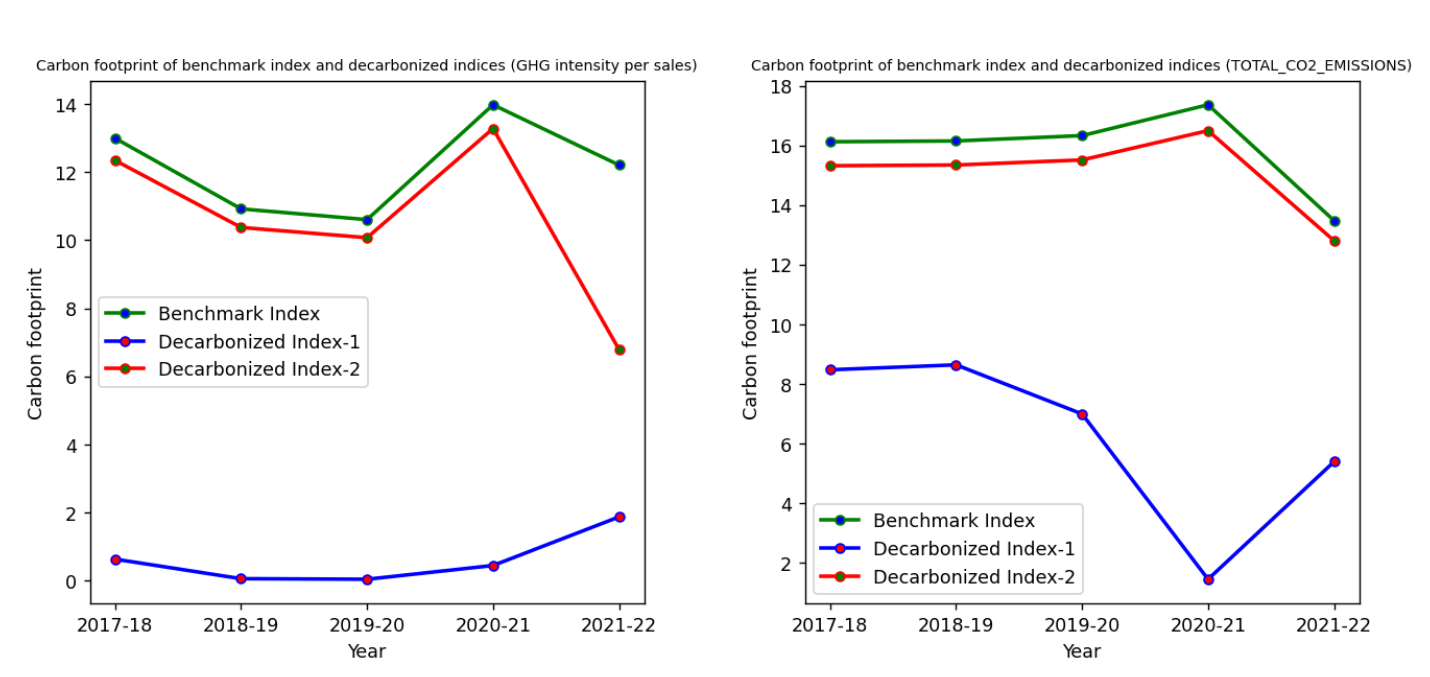}
  \caption{Comparison of carbon footprints of the considered NIFTY-50 and the DIs}
\label{Fig2}
\end{figure}

\begin{figure}[htbp]
\centering
  \centering
  \includegraphics[width=0.8\textwidth,keepaspectratio]{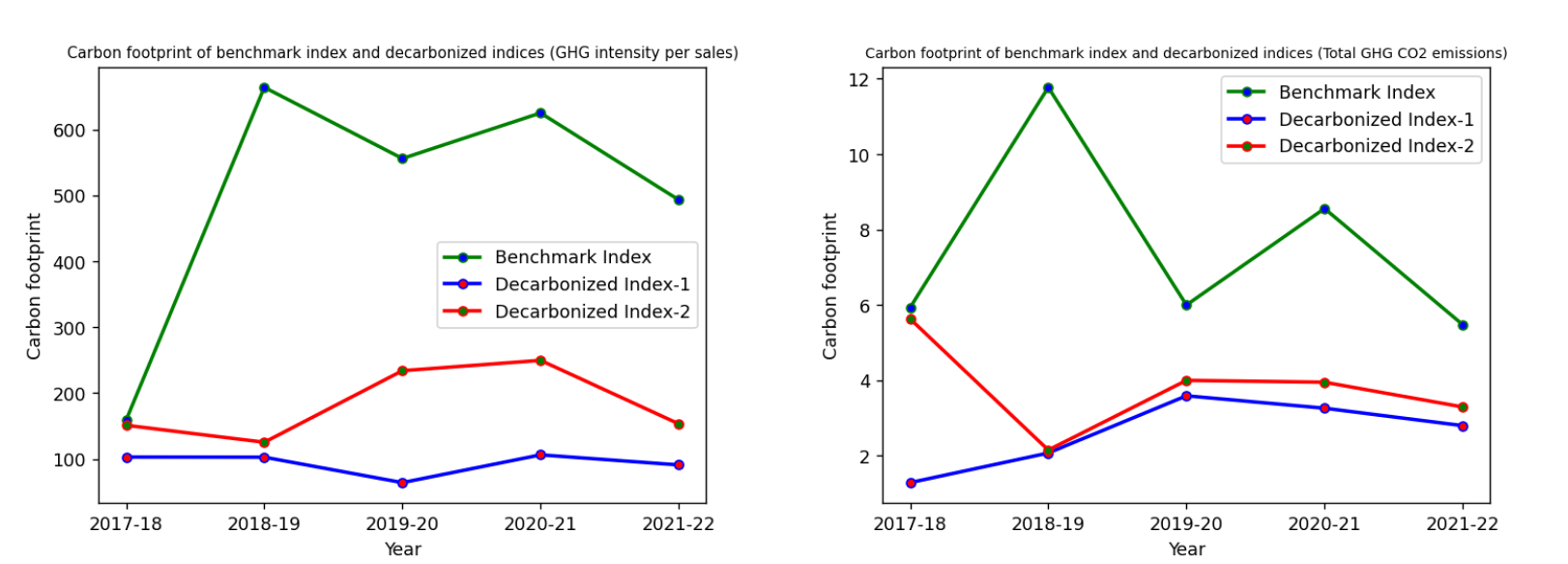}
  \caption{Comparison of carbon footprints of the considered S \& P-500 and the DIs}
\label{Fig3}
\end{figure}

As evident, we observe that the carbon footprints of the DIs are lower compared to the benchmark indices from \Cref{Fig2} and \Cref{Fig3}. A substantial reduction in the carbon footprint of the DIs is observed in all the cases.

Now we look into the in-sample estimation of VaR for the four DIs constructed using a moving window of one year. The VaR estimates of the DIs at $p=0.95$ are reported in \Cref{T2} and \Cref{T3}. The in-sample estimations indicate substantial reductions in carbon footprint for both methods, accompanied by low VaR estimates relative to the benchmark. Both DI\_1 and DI\_2 have VaR estimates close to each other. However, under some scenarios, VaR estimates of DI\_2 are higher than DI\_1. \\

\begin{table}[h]
    \centering
    \caption{VaR estimates of the benchmark portfolio (BP) and DIs at $p=0.95$ for NIFTY-50.}\label{T2}
    \begin{tabular}{|l|ccc|ccc|}
    \hline
    & \multicolumn{3}{|c|}{GHG} & \multicolumn{3}{|c|}{CO2} \\
    Period & BP &  VaR(DI\_1) & VaR(DI\_2) & BP &  VaR(DI\_1) & VaR(DI\_2) \\
     \hline
    2017-18 & 41.735 & 0.436 & 0.599 & 110.801 & 0.840 & 0.643   \\
    2018-19 & 101.062 & 0.523 & 1.067 & 33.697 & 1.502 & 1.161   \\
    2019-20 & 197.143 & 0.626 & 1.333 & 666.889 & 1.819 & 1.392   \\
    2020-21 & 156.305 & 1.189 & 1.562 & 131.275 & 1.162 & 1.527   \\
    2021-22 & 182.445 & 3.107 & 3.107 & 159.810 & 2.918 & 2.841  \\
    \hline
    \end{tabular}
\end{table}

\begin{table}[htbp]
    \centering
    \caption{VaR estimates of the BP and DIs at $p=0.95$ for the S \& P-500.}\label{T3}
    \begin{tabular}{|l|ccc|ccc|}
    \hline
    & \multicolumn{3}{|c|}{GHG} & \multicolumn{3}{|c|}{CO2} \\
    Period & BP & VaR(DI\_1) & VaR(DI\_2) & &  VaR(DI\_1) & VaR(DI\_2) \\
     \hline
    2017-18 & 0.387 & -0.021 & -0.021 & 0.385 & -0.021 & -0.022   \\
    2018-19 & 0.476 & -0.011 & -0.012 & 0.559 & -0.012 & -0.012   \\
    2019-20 & 1.936 & -0.002 & -0.005 & 3.429 & -0.004 & -0.005   \\
    2020-21 & 1.500 & 0.038 & 0.037 & 1.835 & 0.037 & 0.037   \\
    2021-22 & 1.512 & 0.006 & 0.007 & 2.876 & 0.006 & 0.007  \\
    \hline
    \end{tabular}
\end{table}

\subsection{DIs through ES assessment}

\begin{figure}[h]
\centering
  \centering
  \includegraphics[width=0.8\textwidth,keepaspectratio]{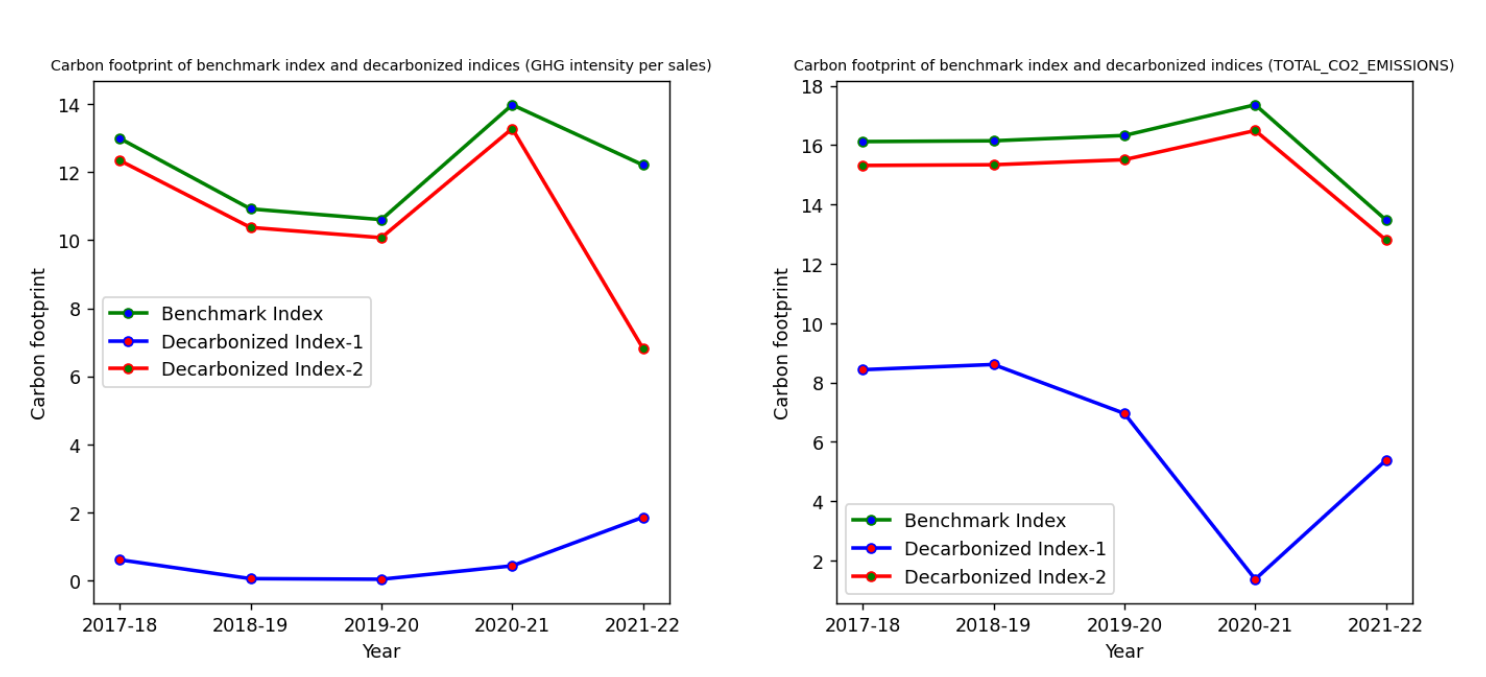}
  \caption{Comparison of carbon footprints of the considered NIFTY-50 and the DIs}
\label{Fig5}
\end{figure}

\begin{figure}[htbp]
\centering
  \centering
  \includegraphics[width=0.8\textwidth,keepaspectratio]{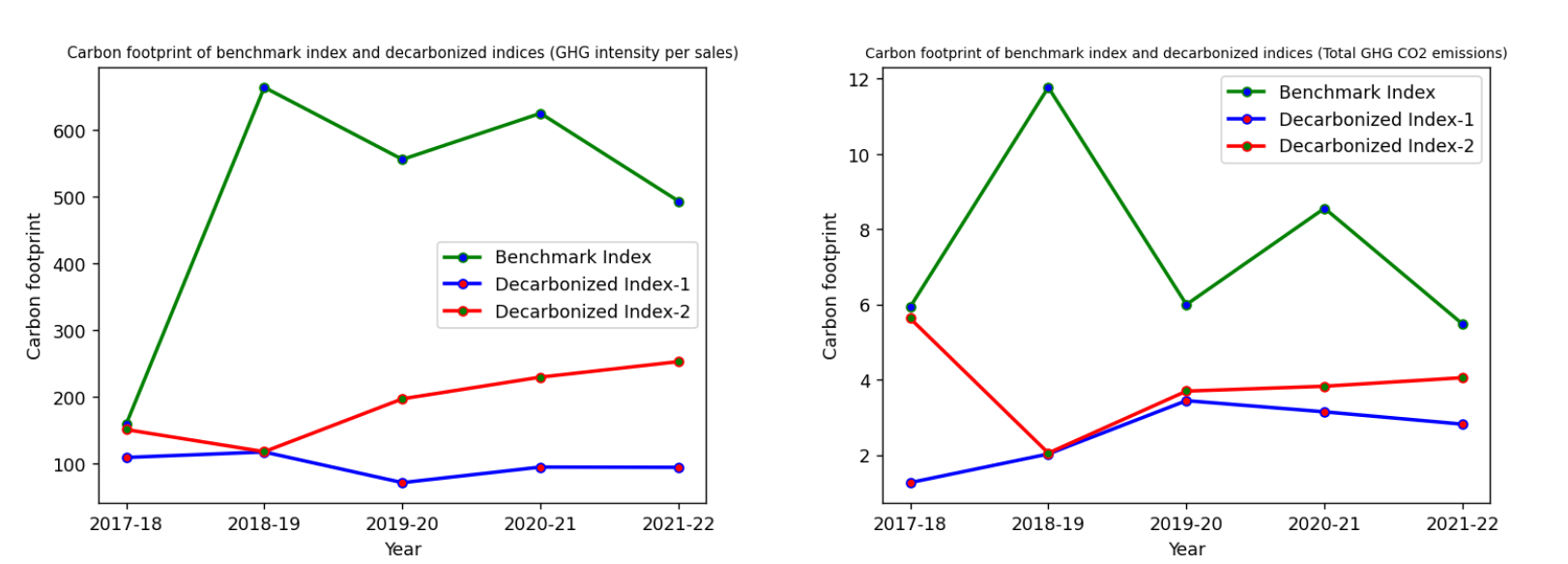}
  \caption{Comparison of carbon footprints of the considered S \& P-500 and the DIs}
\label{Fig6}
\end{figure}

Similarly, in \Cref{Fig5} and \Cref{Fig6}, we compare the carbon footprints of the decarbonized indices with the considered benchmark in each case. Similar outcomes can be observed with substantial reduction in the carbon footprint of the DIs in all the cases.

We will now examine the in-sample estimation of ES for the four DIs developed through a one year moving window. The ES estimates of the DIs at $p=0.95$ are reported in \Cref{T5} and \Cref{T6}. These in-sample estimations reveal significant carbon footprint reductions in both methods, with low ES estimates compared to its benchmark. Both DI\_1 and DI\_2 have ES estimates close to each other. However, under some scenarios, ES values of DI\_2 are higher than DI\_1. Similar observations can also be seen in the case of VaR assessment.\\

\begin{table}[h]
    \centering
    \caption{ES estimates of the BP and DIs at $p=0.95$ for NIFTY-50.}\label{T5}
    \begin{tabular}{|l|ccc|ccc|}
    \hline
    & \multicolumn{3}{|c|}{GHG} & \multicolumn{3}{|c|}{CO2} \\
    Period & BP & ES(DI\_1) & ES(DI\_2) & BP &  ES(DI\_1) & ES(DI\_2) \\
     \hline
    2017-18 & 52.116 & 1.053 & 0.805 & 138.884 & 0.543 & 0.750   \\
    2018-19 & 126.495 & 1.881 & 1.454 & 42.199 & 0.664 & 1.338   \\
    2019-20 & 247.4026 & 2.338 & 1.795 & 836.774 & 0.816 & 1.722   \\
    2020-21 & 195.351 & 1.420 & 1.899 & 163.990 & 1.446 & 1.915   \\
    2021-22 & 228.522 & 3.650 & 3.552 & 200.213 & 3.866 & 3.866  \\
    \hline
    \end{tabular}
\end{table}

\begin{table}[htbp]
    \centering
    \caption{ES estimates of the BP and DIs at $p=0.95$ for S \& P-500.}\label{T6}
    \begin{tabular}{|l|ccc|ccc|}
    \hline
    & \multicolumn{3}{|c|}{GHG} & \multicolumn{3}{|c|}{CO2} \\
    Period & BP &  ES(DI\_1) & ES(DI\_2) &BP &  ES(DI\_1) & ES(DI\_2) \\
     \hline
    2017-18 & 0.472 & -0.017 & -0.017 & 0.469 & -0.017 & -0.018   \\
    2018-19 & 0.589 & -0.009 & -0.009 & 0.691 & -0.009 & -0.009   \\
    2019-20 & 2.436 & 0.003 & 0.003 & 4.304 & 0.003 & 0.003   \\
    2020-21 & 1.826 & 0.042 & 0.042 & 2.251 & 0.043 & 0.042   \\
    2021-22 &1.885 & 0.011 & 0.011 & 3.577 & 0.011 & 0.011  \\
    \hline
    \end{tabular}
\end{table}
\clearpage
\subsection{Out-of-sample calculation}

In the out-of-sample performance, monthly returns are computed for 2018-19 to 2022-23 using weights $\bm W^d$ generated from in-sample calculations conducted in the previous year. The monthly returns of a stock $j$ are estimated using the formula,
\begin{equation*}
r_j=\frac{(p_{jfinal}-p_{jinitial})100}{p_{jinitial}}.
\end{equation*}
Therefore, the monthly returns of a portfolio consisting of $N$ stocks are given by,
\begin{equation}\label{Rt}
R_t=\frac{(WP_{final}-WP_{initial})100}{WP_{initial}},
\end{equation}
where $R_t$ is the percentage return on the portfolio in month $t$, $W$ is the $N\times1$ vector of the portfolio weights, $P_{initial}$ is the vector of stock closing prices at the beginning of the month $t$, and $P_{final}$ is the vector of stock closing prices at the end of the month $t$. The monthly returns of the newly constructed decarbonized portfolios and the considered benchmark portfolios were estimated using equation \eqref{Rt} and are plotted in \Cref{Fig9}, \Cref{Fig10}, \Cref{Fig11}, and \Cref{Fig12}. We observe that the constructed indices track the considered benchmark closely and, on average, outperform the benchmark index.\

We then explore whether, during climate events, the DIs exhibit superior performance compared to their parent benchmark indices. To investigate this effect, we identify and highlight significant climate events from the past few years in the out-of-sample results of our indices. In 2018-2019, COP 24 was held in the month of December, the climate change conference (CCC) in Bonn and Bangkok was held in the months of May and September, and Hurricane Michael caused widespread damage in the state of Florida in October. In 2019-2020, Bonn CCC was held in the month of June, COP 25 was held in December, and the UN summit was held in September. In 2020-2021, Amphan, a devastating tropical cyclone that devastated Bangladesh and Eastern India, particularly West Bengal and Odisha, occurred in May, and the momentum of climate change was held in June. In 2021-2022, UN CCC was held in June, Hurricane Ida caused widespread damage in the U.S. state of Louisiana in August, and COP 26 was in the month of November. In 2022-2023, BON CCC was held in June, hurricanes Ian and Nicole caused widespread damage in the state of Florida in September and November and COP 27 was held in November. Following are the observations from \Cref{Fig9}, \Cref{Fig10}, \Cref{Fig11}, and \Cref{Fig12}.\

\begin{enumerate}
\item NIFTY-50
\begin{itemize}
\item In light of the aforementioned notable climate events, we note that both DIs exceed the standard regarding out-of-sample returns in at least five of the twelve events.
\item DI\_1 of the GHG outperforms the benchmark in 75\% of the events.
\item DI\_2 of the GHG and CO2 outperforms the benchmark in 67\% of the events.
\item DI\_1 of the CO2 outperforms the benchmark in 42\% of the events.
\end{itemize}
\item S\&P-500
\begin{itemize}
    \item In light of the significant climate events discussed above, it is evident that both DIs exceed the benchmark regarding out-of-sample returns in at least 8 of the fifteen events.
    \item DI\_1 of the GHG outperforms the benchmark in 53.4\% of the events.
    \item DI\_2 of GHG and DI\_1 and DI\_2 of CO2 outperform the benchmark in 60\% of the events.
\end{itemize}
\end{enumerate}

\section{Summary and Discussion}

\subsection{Summary of the results}

Our work basically focuses on the theoretical framework and ideas proposed in the construction of DIs by \citep{andersson2016hedging,lakshmi23}. \cite{andersson2016hedging} constructs the DIs for the S \& P-500 and the MSCI Europe benchmark indices, whereas \cite{lakshmi23} constructs DIs for the NIFTY-50 benchmark index. Both these papers involve construction of DIs by minimizing the tracking error. Though tracking error is an essential tool to identify how closely a portfolio follows its benchmark index. However, tracking errors do not give any idea of the risk of the portfolio. Also, we know that the optimal portfolio choice theory by \cite{Mark1952} is based on the mean variance. However, variance takes into account all the deviations from the mean, higher as well as lower, failing to fully address the issue of the right tail of the returns, which is a common concern. Hence, variance is unsuitable as a measure of risk for financial returns. From the literature it is evident that portfolio optimization with respect to the tracking error and the variance is not enough to mitigate the climate risk; we do need the measures of downside risks. In this paper we construct the DIs by minimizing the mean-VaR and mean-ES, which means an investor possessing a DI is protected against the timing risk associated with climate mitigation policies, which are anticipated to adversely affect companies with high carbon footprints. This is due to the design of the DIs, which aim to sustain a low VaR and ES in relation to the benchmark indices. We construct two DIs from the established benchmarks and demonstrate their efficacy for the Indian (NIFTY-50 index) and the U.S. (S \& P-500 index) economies.\

The DIs obtained through VaR and ES assessment show a substantial reduction in the carbon footprint of the DIs in all the cases as compared to their benchmark indices, NIFTY-50 and S \& P-500. Specifically, we show that the resulting index significantly lowers total carbon impact, acting as a hedge against climate risks. In-sample calculations reveal that the VaR and ES estimated at $p=0.95$ of the DIs are very low compared to their benchmark indices. We also observe that the VaR and ES estimates of both the DIs (GHG and CO2) are very close to each other, except for a few instances where the VaR and ES of DI\_2 are higher than DI\_1. On the other hand, out-of-sample results demonstrate that both indices (GHG and CO2) outperform the benchmark indices during major climate events throughout the five years. Specifically, in the NIFTY-50, DI\_1 of the GHG outperforms the benchmark in 75\% of the events and DI\_2 of the GHG and DI\_1 and DI\_2 of CO2 outperform the benchmark in 60\% of the events. These results show that the DIs relation to the climate events is quite impressive in both the NIFTY-50 and S \& P-500. \

Both Cyclone Amphan, which resulted in US\$14 billion in damages and 103 fatalities in India, and Hurricane Michael, which resulted in US\$30 billion in damages and at least 59 fatalities in the U.S., caused severe destruction. Typically, these factors lead to an increase in relative risk. And it is well-known that a higher relative risk for a specific location is associated with an increased combination of hazard, exposure, and vulnerability. Yet, under these scenarios we observe that our proposed DIs outperform the benchmark.\

\subsection{Implications from our study}

In the absence of climate mitigation initiatives, these low-carbon indices will yield returns that are generally comparable to the benchmark, occasionally underperforming or outperforming it. During significant climate events, the DIs have surpassed their benchmark. Upon the imposition or anticipation of a carbon penalty, these indices will surpass the actual benchmark in nearly all instances. However, the results indicate that the DIs perform adequately, suggesting that these investors will not incur a significant carbon penalty at this time. However, these distinctions can significantly assist policymakers in estimating carbon pricing. In the most adverse scenario, if carbon pricing is not implemented in the future, investing in low-carbon portfolios may still not result in significant losses for investors. The clear communication of the excluded and included stocks will reward the included companies and also help create a sense of awareness and healthy competition among the excluded companies to get back in the index by reducing their emissions. These indices will also help improve scientific quantification, accounting, and reporting of
emissions by companies.

\subsection{Limitations}

The quantification of company-specific GHG emissions to ascertain the carbon footprint is crucial in the formulation of DIs. A significant problem we encountered was locating dependable carbon data from open sources. Another significant issue was the substantial amount of missing data in the available sources, attributable to inadequate accounting and reporting by the companies. We found it necessary to exclude certain valuable stocks from the benchmark index owing to the lack of available carbon data. The stocks that were not included during the construction phase may represent the more volatile elements of the benchmark index, exhibiting heightened sensitivity to climate change and related policies. Another concern is whether the GHG emissions are accurately measured and reported in accordance with the GHG protocol and if there might be an inherent bias in the measurement process.\

The regression results indicate that the Fama-French five variables for the American stock market and the four common components for the Indian market inadequately explain stock returns, which are mostly influenced by stock-specific risks. However, we choose this approach to streamline the issue and minimize errors associated with extensive variance-covariance matrices. And also due to the availability of these factors data in open sources. A further drawback was the unavailability of historical market capitalization data, resulting in the assumption that the weights of the benchmark indices remained constant over the years. The market cap data also depends on the time at which the data was downloaded, as it keeps changing. So slight variations can be observed in the market cap data due to the difference in the timings of data access. We have not followed a sector-by-sector filtering approach for construction of the DIs, due to which the newly constructed indices might have slightly different sector compositions from their parent benchmark index, especially when constructed using Method 1, where we drop $k$ stocks. We can extend this to factoring in sector compositions as well for better results.





\bibliography{refs}

\appendix
\section{Figures}

\begin{figure}[h]
\centering
  \centering
\includegraphics[width=0.6\textwidth,keepaspectratio]{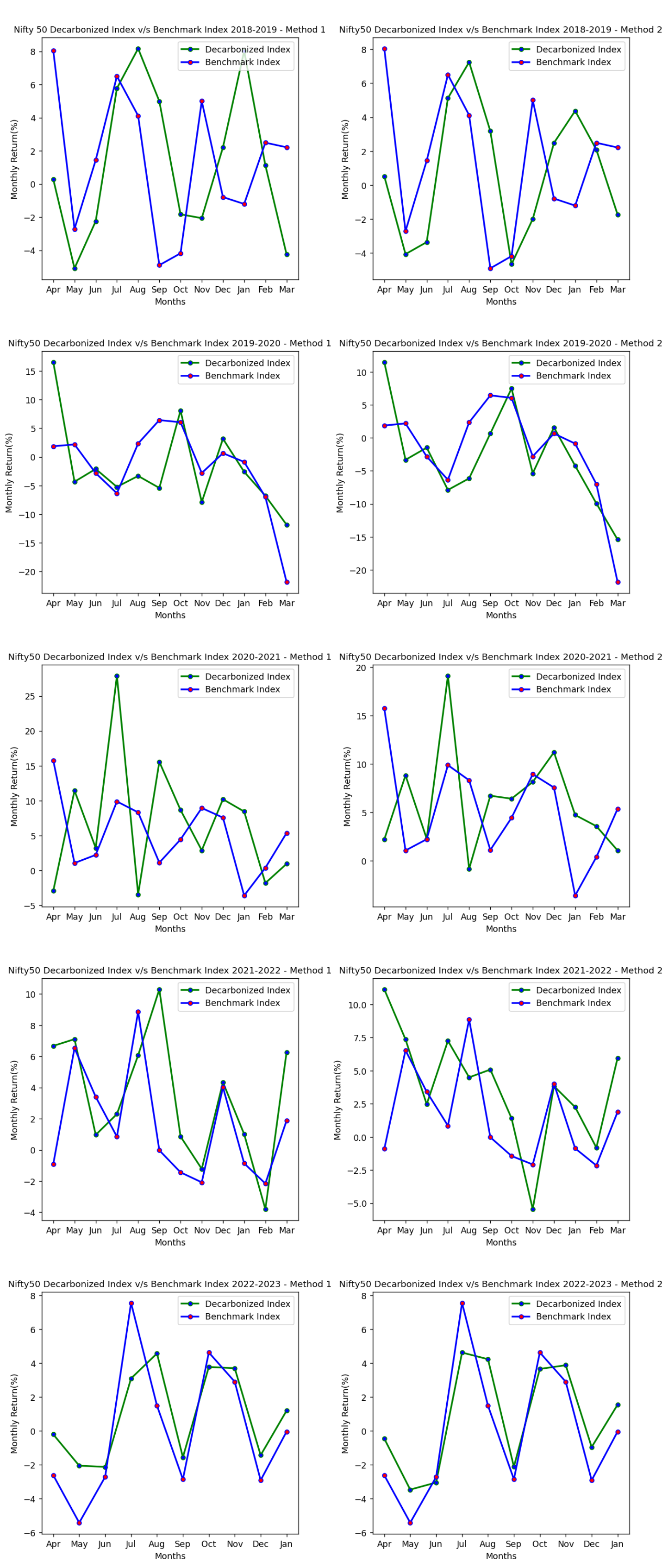}
  \caption{Monthly returns of DIs vs Benchmark index for NIFTY-50 (GHG)}
\label{Fig9}
\end{figure}

\begin{figure}[htbp]
\centering
  \centering
\includegraphics[width=0.6\textwidth,keepaspectratio]{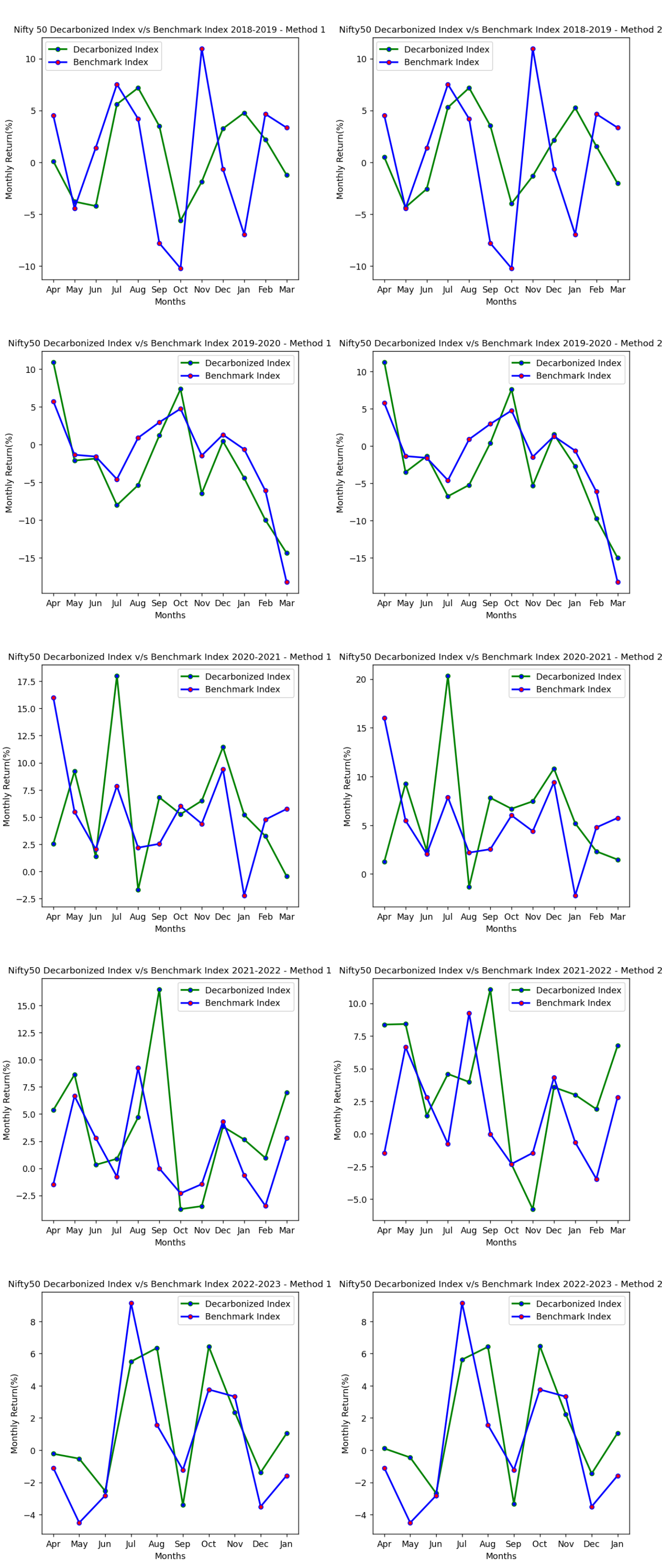}
  \caption{Monthly returns of DIs vs Benchmark index for NIFTY-50 (CO2)}
\label{Fig10}
\end{figure}

\begin{figure}[htbp]
\centering
  \centering
\includegraphics[width=0.7\textwidth,keepaspectratio]{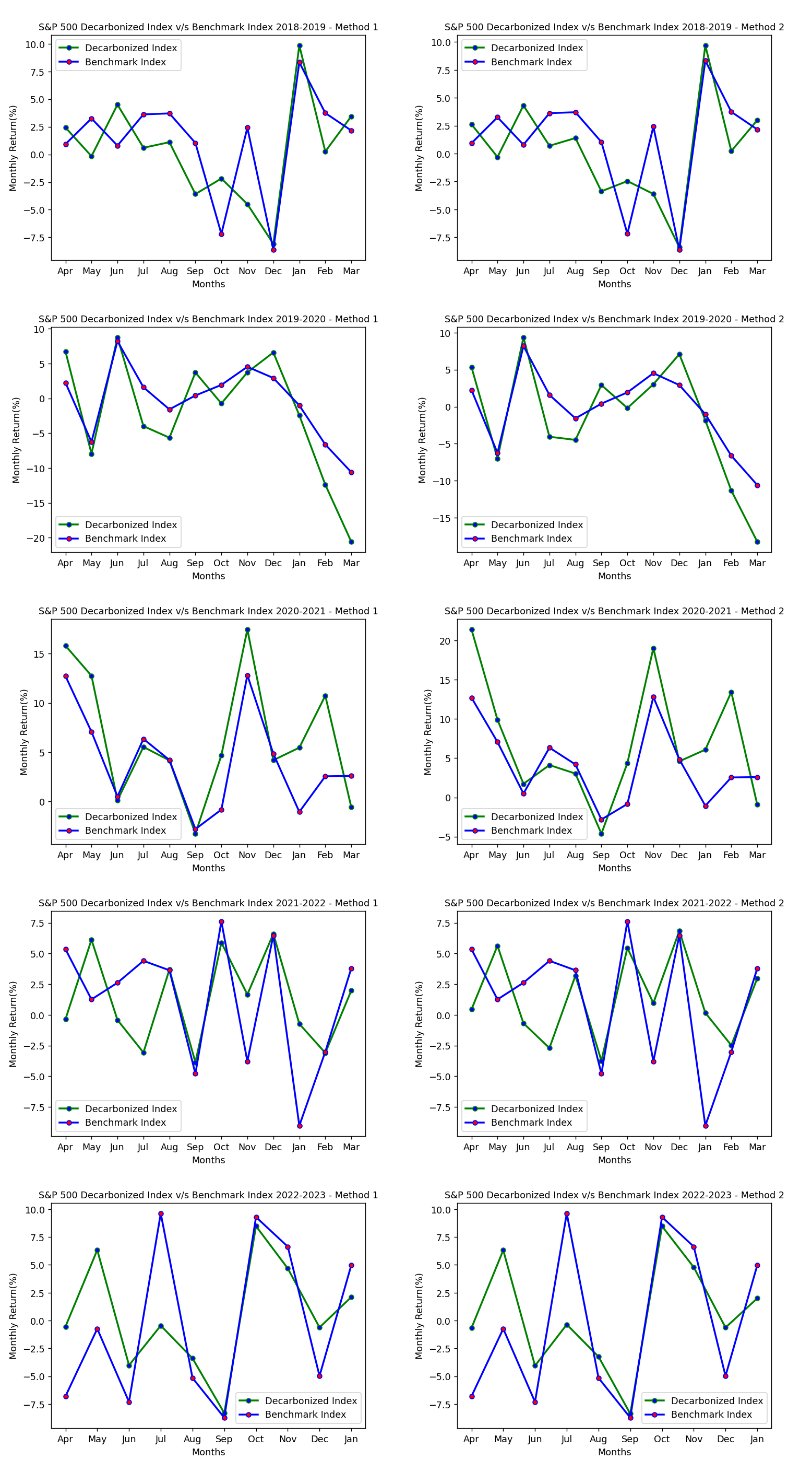}
  \caption{Monthly returns of DIs vs Benchmark index for S\&P-500 (GHG)}
\label{Fig11}
\end{figure}

\begin{figure}[htbp]
\centering
  \centering
\includegraphics[width=0.7\textwidth,keepaspectratio]{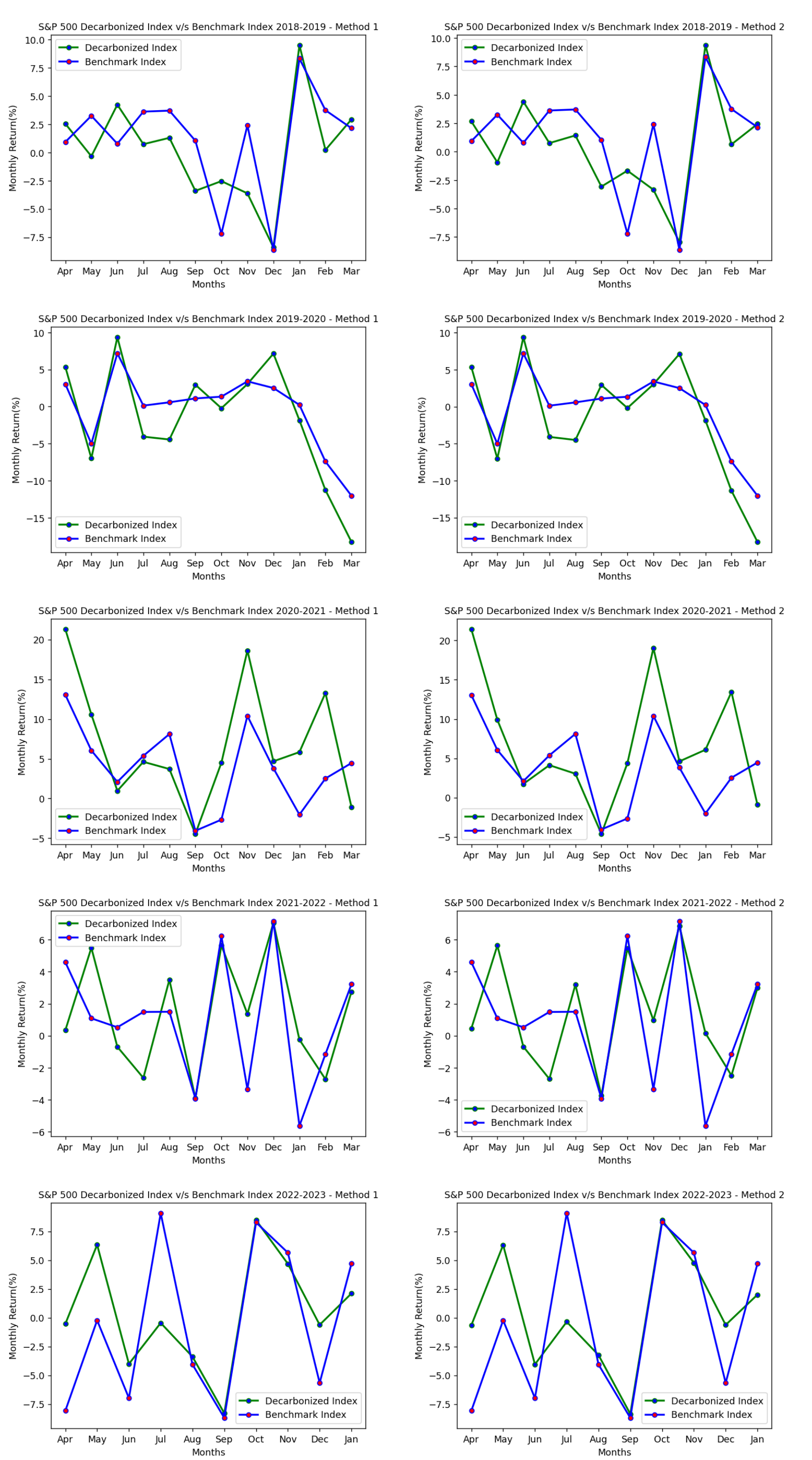}
  \caption{Monthly returns of DIs vs Benchmark index for S\&P-500 (CO2)}
\label{Fig12}
\end{figure}

\end{document}